\PassOptionsToPackage{final}{hyperref}

\documentclass[sigconf]{acmart}

\makeatletter                   
\def\mdseries@tt{m}             
\makeatother                    

\usepackage{adjustbox}
\usepackage{bookmark}
\usepackage{booktabs}
\usepackage{caption}
\usepackage{cleveref}
\usepackage{enumitem}
\usepackage{filecontents}
\usepackage{flushend}
\usepackage[T1]{fontenc}
\usepackage[final]{listings}
\usepackage{makecell}
\usepackage{mathpartir}
\usepackage[framemethod=tikz]{mdframed}
\usepackage[frozencache]{minted}
\usepackage{savesym}
\usepackage{soul}
\usepackage{subcaption}
\usepackage{tikz}
\usepackage{xspace}

\usetikzlibrary{calc}

\makeatletter
\@namedef{T1/zi4/m/it}{<->ssub*zi4/m/n}
\makeatother

\newmdenv[
    nobreak,
    middlelinewidth=.8pt,
    frametitlefont=\bfseries,
    leftmargin=.3cm,
    rightmargin=.3cm,
    skipabove=\topsep,
    skipbelow=\topsep,
    singleextra={
        \draw[line width=1.6pt,white,] ($(O|-P)+(.2cm,0)$) -- ($(P)-(.2cm,0)$); 
        \draw[line width=1.6pt,white,] ($(O)+(.2cm,0)$) -- ($(P|-O)-(.2cm,0)$); 
    },%
]{standout}

\savesymbol{comment}
\usepackage[final,authormarkup=none]{changes}

\crefformat{section}{\S#2#1#3}
\crefformat{sections}{\S\S#2#1#3}
\crefname{figure}{Fig.}{Figs.}
\Crefname{figure}{Fig.}{Figs.}

\copyrightyear{2020}
\acmYear{2020}
\setcopyright{acmcopyright}
\acmConference[MSR '20]{17th International Conference on Mining Software Repositories}{October 5--6, 2020}{Seoul, Republic of Korea}
\acmBooktitle{17th International Conference on Mining Software Repositories (MSR '20), October 5--6, 2020, Seoul, Republic of Korea}
\acmPrice{15.00}
\acmDOI{10.1145/3379597.3387498}
\acmISBN{978-1-4503-7517-7/20/05} 


\begin{document}

\restoresymbol{change}{comment}

\definechangesauthor[name=Jordan, color=blue]{J}
\definechangesauthor[name=Chris, color=red]{C}
\definechangesauthor[name=Tom, color=green]{T}

\title{A Dataset of Dockerfiles}

\newcommand{\UWMad}[1][ersity]{Univ#1 of Wisconsin--Madison} 

\author{Jordan Henkel}
\orcid{0000-0003-3862-249X} 
\affiliation[obeypunctuation=true]{%
  \institution{\UWMad},
  \country{USA}
}
\email{jjhenkel@cs.wisc.edu}

\author{Christian Bird}
\affiliation[obeypunctuation=true]{%
    \institution{Microsoft Research},
    \country{USA}
}
\email{Christian.Bird@microsoft.com}

\author{Shuvendu K. Lahiri}
\affiliation[obeypunctuation=true]{%
  \institution{Microsoft Research},
  \country{USA}
}
\email{Shuvendu.Lahiri@microsoft.com}

\author{Thomas Reps}
\orcid{0000-0002-5676-9949} 
\affiliation[obeypunctuation=true]{%
  \institution{\UWMad},
  \country{USA}
}
\email{reps@cs.wisc.edu}

\captionsetup{labelfont={sf, small},textfont={sf, small}}

\renewcommand{\shortauthors}{J. Henkel, C. Bird, S. Lahiri, and T. Reps}

\begin{abstract}
  Dockerfiles are one of the most prevalent kinds of DevOps artifacts
  used in industry. Despite their prevalence, there is a lack of 
  sophisticated semantics-aware static analysis of Dockerfiles. In this paper,
  we introduce a dataset of approximately 178,000 unique Dockerfiles collected from GitHub.
  To enhance the usability of this data, we describe five representations we have
  devised for working with, mining from, and analyzing these Dockerfiles.
  Each Dockerfile representation builds upon the previous ones, and the final representation,
  created by three levels of nested parsing and abstraction, makes tasks
  such as mining and static checking tractable. The Dockerfiles, in each of the
  five representations, along with metadata and the tools used to shepard the data from
  one representation to the next are all available at: https://doi.org/10.5281/zenodo.3628771. 
\end{abstract}

\maketitle


\section{Introduction}

DevOps artifacts in general, and Dockerfiles in particular, represent a relatively
under-served area with respect to advanced tooling for assisting developers. \added[id=J]{We focus on Docker because it is the most prevalent DevOps artifact in industry (some 79\% of IT companies use it~\citep{portworx}) and the de-facto container technology in OSS~\citep{CitoEmpiricalDockerEcoMSR2017,zhang2018cd}. Nevertheless,}
the VS Code Docker extension, with its over 3.7 million unique installations, features relatively shallow syntactic support \citep{vscode:Docker}. One possible
reason for the lack of advanced tooling may be the challenge of \emph{nested languages}.
Many DevOps artifacts have relatively simple top-level structure---YAML and JSON are
two popular top-level choices---although some tools, like Docker, have a custom top-level
language. Oftentimes some form of embedded scripting
language (primarily Bash) is nested within the top-level syntax. Furthermore, within an embedded Bash script, there are
any number of user-authored or distribution-provided scripts and packages. Each
of these tools, in turn, induce new sub-languages based on their grammar of options, arguments,
and inputs. (As a simple example, think of Unix utilities like \verb|awk|, \verb|sed|, and \verb|grep|.)

These third-level sub-languages represent a road-block
to a wholistic understanding of many DevOps artifacts. Even advanced tools,
such as Hadolint \citep{github:Hadolint}, make no attempt to parse further than the second-level
of embedded shell code. The lack of structured representations at this third-level
of embedded languages is a major hindrance to both mining and static checking of
Dockerfiles and DevOps artifacts, in general \added[id=J]{\citep{SidhuLackOfReuseTravisSaner2019}.}

With the dataset of Dockerfiles described in this paper, we make the following
core contribution:

\begin{standout}
\added[id=J]{Abstract Syntax Trees (ASTs)} for a set of 178,000 unique Dockerfiles with \emph{structured representations} of the
(i) top-level syntax, (ii) second-level embedded shell, and (iii) third-level
options and arguments for the 50 most commonly used utilities, and the tools used
to generate each of these representations.
\end{standout}

\vspace{-0.4cm}
\section{Dockerfile Collection}\label{Se:Collection}

To capture a sufficiently large set of Dockerfiles, we made use
of GitHub's API to query for repository metadata. To begin with,
we downloaded metadata for every public repository with ten or more stars
from January 1st, 2007 to June 1st, 2019. This process yielded approximately
900,000 metadata entries (each corresponding to one repository).

With repository metadata in hand, we began the next phase of data collection. 
For each of the 900,000 repository metadata entries, we again used GitHub's API
to select a recursive listing of all the files and directories present in each
repository. We stored this data, along with the repository metadata entries, in
a relational database. Note that, at this point, we have avoided downloading
repositories directly (via a fetch or clone). This approach avoids the problem of storing
an inordinate amount of data (most of which we are uninterested in).

Next, we ran a case-insensitive query against our database to find all files in all repositories with
names containing the string \verb|dockerfile|.
This process yielded approximately 250,000 matches. At this point, we began to
download each likely Dockerfile from GitHub individually. As files were downloaded, they
were saved to disk. In the event of a failed download request, the download was
re-tried up to five times before skipping the errant file.

Finally, we applied a Dockerfile parser from the \verb|dockerfile| Python package \citep{github:dockerfile}.
We performed this step to reduce the number of non-Dockerfile files that may have been present
due to our very basic initial filtering. Files that failed to parse were simply deleted.
After this process, we were left with approximately 219,000 Dockerfiles.

\vspace{-0.25cm}
\subsection*{Gold Files}

Within the set of Dockerfiles we collected, there are 432 Dockerfiles from the
\verb|docker-library/| organization on GitHub. These files are of particular interest
because they come from repositories managed and maintained by Docker experts, and are,
presumably, exemplars of high-quality Dockerfile writing. For convenience, we have
duplicated these files and stored them, alongside the full corpus, for each representation
we describe in \cref{Se:Reprs}. In our artifact, the Gold files follow the naming convention \verb|gold.*| whereas the
overall corpus follows the convention \verb|github.*|.

\vspace{-0.25cm}
\subsection*{Metadata}

In addition to the source-level Dockerfiles we obtained, we also captured metadata corresponding to each Dockerfile.
This metadata captures information such as the repository from which the Dockerfile was originally downloaded, the time
of the original download, the sub-directory in which the Dockerfile originally resided, and various other ancillary details.
For completeness, we provide this metadata in the \verb|./datasets/5-dockerfile-metadata|
directory of our artifact. An example of accessing this data is provided below.

\medskip\noindent\textbf{Example Usage:}

\begin{minted}[fontsize=\footnotesize]{bash}
cat ./5-dockerfile-metadata/github.jsonl.xz \
  | xz -cd | grep 'file_id":133495483' | jq
\end{minted}

\noindent Running the above should produce:

\begin{minted}[fontsize=\footnotesize]{json}
{
  "file_id": 133495483,
  "file_sha": "a2f4e76c9a16dbdaecf623f2878dd66b9609c371",
  "file_url": "https://github.com/.../blob/master/Dockerfile",
  "repo_branch": "master",
  "repo_full_name": "dordnung/System2",
  ...
}
\end{minted}

\vspace{-0.25cm}
\section{Dockerfile Representations}%
\label{Se:Reprs}

We now present details about the various representations of this data. The Dockerfiles, at the source
level, are of limited use in structured tasks like mining and static checking. To
provide more readily usable data we transformed the original Dockerfiles into several
representations, each building upon the last, resulting in, ultimately, rich
Abstract Syntax Trees (ASTs) on which pattern mining and static checking are tractable.

\vspace{-0.25cm}
\subsection*{Representation 0: Source Files}

In the first representation, we created a compressed tar archive of the directory
of Dockerfiles we originally collected. We did the same for the subset of Gold files. These
compressed tar archives are present in the \verb|./datasets/0a-original-dockerfile-sources| directory
of our artifact.

\medskip\noindent\textbf{Example Usage:}

\begin{minted}[fontsize=\footnotesize]{bash}
tar -xvJf ./0a-original-dockerfile-sources/github.tar.xz
cd ./sources
cat 484097305.Dockerfile
\end{minted}

\noindent Running the above should produce:

\begin{minted}[fontsize=\footnotesize]{docker}
FROM busybox
EXPOSE 80/tcp
COPY httpserver .
CMD ["./httpserver"]
\end{minted}

\subsection*{Representation 1: De-duplicated Source Files}

One common issue in datasets sourced from GitHub is duplication. For DevOps artifacts, this issue
is compounded by the common tactics of finding a workable artifact from another similar repository,
or using one of many ``catch-all'' patterns. In either case, duplicate files may likely be created.
To address duplication, we removed files from \emph{Representation 0} that were non-unique based on a SHA 256
hash (calculated using \verb|sha256sum|). We then generated compressed tar archives as before. 
These archives are present in the \verb|./datasets/0b-deduplicated-dockerfile-sources| directory
of our artifact.

\medskip\noindent\textbf{Example Usage:}

\begin{minted}[fontsize=\footnotesize]{bash}
tar -xvJf ./0b-deduplicated-dockerfile-sources/github.tar.xz
cd ./deduplicated-sources
cat f9f9726d2643993eb2176491858b7875ae332d05.Dockerfile 
\end{minted}

\noindent Running the above should produce:

\begin{minted}[fontsize=\footnotesize]{docker}
# https://hub.docker.com/r/consensysllc/go-ipfs/
# THANKS!!!!!

FROM ipfs/go-ipfs
COPY start_ipfs.sh /usr/local/bin/start_ipfs
\end{minted}

\subsection*{Representation 2: Phase-I ASTs}

In the next representation, we make the transition from source-level Dockerfiles to an encoding
of Abstract Syntax Trees for Dockerfiles. We applied the parser from Python's \verb|dockerfile|
package to obtain a Concrete Syntax Tree (CST). We then applied significant post-processing
to obtain something closer to an AST\@. Additionally, we checked to make sure the
directives extracted by the \verb|dockerfile| package were actually known directives (due to this package's 
permissive parser, a small number of invalid files manage to generate valid parse trees---we detected and
rejected these files at this stage). We encoded the whole corpus (and the Gold subset) via compressed
JSON lines files (JSONL). A JSONL file stores, on each line, one valid JSON object representing a single entity.
These JSONL files are present in the \verb|./datasets/1-phase-1-asts| directory of our artifact. 

\medskip\noindent\textbf{Example Usage:}

\begin{minted}[fontsize=\footnotesize]{bash}
cat ./1-phase-1-dockerfile-asts/github.jsonl.xz \
  | xz -cd \
  | grep '3d0d691c1745e14be0f1facd14c49e3fbbb750d8' \
  | jq
\end{minted}

\noindent Running the above should produce:

\begin{minted}[fontsize=\footnotesize]{json}
{
  "type": "DOCKER-FILE",
  "children": [{
    "type": "DOCKER-FROM",
    "children": [{
      "type": "DOCKER-IMAGE-NAME",
      "value": "solaris",
      "children": []
    }]
  }, ..., {
    "type": "DOCKER-CMD",
    "children": [{
      "type": "DOCKER-CMD-ARG",
      "value": "./httpserver",
      "children": []
    }]
  }],
  "file_sha": "3d0d691c1745e14be0f1facd14c49e3fbbb750d8"
}
\end{minted}

\subsection*{Representation 3: Phase-II ASTs}

One key insight and contribution we bring to Dockerfile analysis is the necessity of dealing
with the nested languages present in Dockerfiles. The most immediate nested language in a typical
Dockerfile is some form of shell scripting in \verb|RUN| statements. Primarily, these
statements contain valid Bash (but, in principal, scripts for other shells such as Window's Powershell
are permitted). In \emph{Representation 3}, we took the ASTs from \emph{Representation 2} and, for each AST,
identified and parsed any embedded Bash. We assumed that the child of any \verb|RUN| statement contains
embedded Bash, and employed ShellCheck \citep{github:ShellCheck} to parse these literal nodes into sub-trees.
We again stored the results as compressed JSONL files, which can be found in the \verb|./datasets/2-phase-2-dockerfile-asts|
directory of our artifact.

\medskip\noindent\textbf{Example Usage:}

\begin{minted}[fontsize=\footnotesize]{bash}
cat ./2-phase-2-dockerfile-asts/github.jsonl.xz \
  | xz -cd \
  | grep '972b56dc14ff87faddd0c35a5f3b6a32597a36ed' \
  | jq
\end{minted}

\noindent Running the above should produce:

\begin{minted}[fontsize=\footnotesize]{json}
{
  "type": "DOCKER-FILE",
  "file_sha": "972b56dc14ff87faddd0c35a5f3b6a32597a36ed",
  "children": [..., {
    "children": [{
      "children": [{
        "children": [..., {
          "children": [{
            "value": "npm",
            "children": [],
            "type": "BASH-LITERAL"
          }],
          "type": "BASH-COMMAND-COMMAND"
        }, {
          "children": [{
            "value": "install",
            "children": [],
            "type": "BASH-LITERAL"
          }, {
            "value": "--production",
            "children": [],
            "type": "BASH-LITERAL"
          }],
          "type": "BASH-COMMAND-ARGS"
        }],
        "type": "MAYBE-SEMANTIC-COMMAND"
      }],
      "type": "BASH-SCRIPT"
    }],
    "type": "DOCKER-RUN"
  }, ...]
}
\end{minted}

\subsection*{Representation 4: Phase-III ASTs}

Although the previous representation is workable and used
in both Hadolint \citep{github:Hadolint} and recent work on Dockerfiles \citep{CitoEmpiricalDockerEcoMSR2017}, one of the
core contributions of this dataset is a richer representation
of Dockerfiles based on the use of many parsers. First, we created parsers for
each of the 50 most used Bash commands in Dockerfiles. \added[id=J]{(Here, the 50 most used Bash commands were identified, empirically, by counting and ranking the Bash commands present in our Phase-II ASTs.)} Next, to
arrive at \emph{Representation 4}, we took each Phase-II AST
and found every sub-tree (in the embedded Bash that we parsed
as part of Phase-II) that corresponded to one of the 50 most
frequently used Bash commands in our corpus of Dockerfiles. For each of these
corresponding sub-trees, we extracted them and applied
the appropriate parser for the command. The
results of this third-level parse were then used to replace the
removed sub-tree.

The example usage below highlights this process: note how the
\verb|MAYBE-SEMANTIC-COMMAND| node from the previous Phase-II AST
has been replaced by a new \verb|SC-NPM-INSTALL| sub-tree. This
new sub-tree has structured nodes corresponding to the various flags,
options, and parameters defined by the \verb|npm| utility. It is
in this Phase-III representation that we finally have the ability
to mine, in a structured way, patterns such as: ``\verb|npm|'s \verb|--production| flag
must always be present when running the \verb|npm install| sub-command''.

To make this extra level of parsing possible and less onerous,
we leveraged the fact that all of the popular Bash utilities have some
form of embedded help documentation (accessible either through a flag or manual pages).
This documentation often describes, in detail, the schema of allowable flags,
options, and parameters. Unfortunately, these help documents are written in
natural language. Therefore, we wrote a parser generator that takes
structured schemas that are close, in spirit, to help documentation. With this
specially designed input format, it became much easier to write schemas
and generate parsers. In fact, it took us on average between 15 and 30
minutes to encode individual schemas for popular command-line utilities.
\added[id=J]{Encoding schemas, although manual work, is a one-time process---the parsers
we generate are efficient (operating, commonly, in milliseconds) and, once generated,
parsers can be used with any DevOps artifact containing nested Bash, not just Dockerfiles.}

Our Phase-III ASTs are stored as compressed JSONL files. These files reside in
the \verb|./datasets/3-phase-3-dockerfile-asts| directory of our artifact.
Additionally, the schemas we use for parser generation are available in the
\begin{description}
  \item \verb|./datasets/3-phase-3-.../generate/enrich/commands|
\end{description}
 directory. Each
schema is encoded as a YAML file to strike a balance between programmatic ease
of use and human readability. \added[id=J]{These schemas encode both flags with their types (boolean, array, etc.) and the various usage scenarios allowed by a command. Scenarios mostly mirror a command's allowable sub-commands (e.g., \texttt{git clone/add/\ldots}). Each scenario has its own configuration and, via YAML Merge Keys, scenarios may inherit common flag definitions. (This feature is useful for common flags like \texttt{-h/--help}.)}

\medskip\noindent\textbf{Example Usage:}

\begin{minted}[fontsize=\footnotesize]{bash}
cat ./3-phase-3-dockerfile-asts/github.jsonl.xz \
  | xz -cd \
  | grep '972b56dc14ff87faddd0c35a5f3b6a32597a36ed' \
  | jq
\end{minted}

\noindent Running the above should produce:

\begin{minted}[fontsize=\footnotesize]{json}
{
  "file_sha": "972b56dc14ff87faddd0c35a5f3b6a32597a36ed",
  "type": "DOCKER-FILE",
  "children": [..., {
    "children": [{
      "children": [{
        "children": [{
          "children": [], "type": "SC-NPM-F-PRODUCTION"
        }],
        "type": "SC-NPM-INSTALL"
      }],
      "type": "BASH-SCRIPT"
    }],
    "type": "DOCKER-RUN"
  }, ...]
}
\end{minted}

\subsection*{Representation 5: Abstracted Phase-III ASTs}

For our final representation of Dockerfiles, we applied a set
of simple abstractions to each literal value present in our
Phase-III ASTs. Each regular expression is assigned a name, and
when a given expression matches a literal node, a new node is
inserted into the tree as a child of the literal node. The type of the
newly inserted node is set to the name of the matched regular expression.

While these regular expressions are hand-designed,
their purpose is to supplement our ASTs with 
possibly useful information without going so far as to implement
something like a fourth phase of parsing. Furthermore, having
abstractions as our final step of dataset processing introduces
a convenient entry-point for doing simple exploratory analysis. (As an example,
one could add regular expressions to identify GitHub URLs and then, in a
structured pass over Phase-III abstracted ASTs, identify how often
\verb|npm| is used with a GitHub URL as an argument in lieu of a package name).

Our abstracted Phase-III ASTs are stored as compressed JSONL files.
These files reside in the \verb|./datasets/4-abstracted-asts| directory
of our artifact. Additionally, the regular expressions we use for abstraction
are present in the
\begin{description}
  \item \verb|./datasets/4-abstracted-.../generate/abstractions.py|
\end{description}
 file. Each regular expression is encoded, with its name, into a Python file as an array of
 tuples.

\medskip\noindent\textbf{Example Usage:}

\begin{minted}[fontsize=\footnotesize]{bash}
cat ./4-abstracted-asts/github.jsonl.xz \
  | xz -cd \
  | grep 'aaf505fc6efd672143ac63292122207db3f8b19b' \
  | jq
\end{minted}

\noindent Running the above should produce:

\begin{minted}[fontsize=\footnotesize]{json}
{
  "file_sha": "aaf505fc6efd672143ac63292122207db3f8b19b",
  "type": "DOCKER-FILE",
  "children": [..., {
    "children": [{
      "children": [{
        "children": [..., {
          "children": [{
            "children": [{
              "children": [{
                "type": "ABS-PROBABLY-URL",
                "children": []
              }, {
                "type": "ABS-URL-PROTOCOL-HTTPS",
                "children": []
              }],
              "type": "BASH-SINGLE-QUOTED"
            }, ...],
            "type": "BASH-ARRAY"
          }],
          "type": "BASH-ASSIGN-RHS"
        }],
        "type": "BASH-ASSIGN"
      }, ...],
      "type": "BASH-SCRIPT"
    }],
    "type": "DOCKER-RUN"
  }, ...]
}
\end{minted}

\vspace{-0.35cm}
\section{Dataset Usages}

We have used the dataset presented here to carry out a study on the feasibility of automated rule mining from Dockerfiles. In addition, we
have also manually curated a collection of \emph{Gold Rules}, and used these rules to gather general statistics on the
incidence of rule violations in Dockerfiles on GitHub. In that study we found that, on average, there are \emph{five times more} rule
violations in the overall corpus of Dockerfiles compared to the number of violations in the Gold Files introduced in~\cref{Se:Collection}. Moreover, we found
that frequent sub-tree mining \citep{chi2005frequent,chi2005mining}, with the help of some light modifications, can effectively mine
Tree Association Rules \citep{mazuran2009mining} from this corpus. For comprehensive details and analysis, see \citet{henkel2020devops}.

In addition to the Dockerfiles presented earlier, we have also made the Gold Rules available in the \verb|./datasets/6-gold-rules| directory
of our artifact. Each rule is rendered as a simple JavaScript Object, and encoded into a TypeScript file for easy usage in a downstream
application, such as a static rule checker. 

\vspace{-0.17cm}
\section{Future Directions}

Although we successfully implemented an automated rule miner and static-checking engine using this data, our techniques have several
key limitations. First, our automated rule miner is limited in the kind of Tree Association Rules it can mine. Expanding the class
of minable rules would be a significant advance. Second, we have not yet investigated the possibility of using this data
in the context of \emph{repairs}. It is likely that one can to use these more structured representations of Dockerfiles to bootstrap
interesting research on the automated repair of common Docker mistakes. Finally, there are a number of other interesting uses
for this data outside of rule mining, checking, and violation repairs. In particular, encoded within these Dockerfiles is a wealth
of information on the kinds of tools being used in production environments, and, more critically, \emph{the dependencies} among various
pieces of production software. We believe that research in this direction would be of great \added[id=J]{interest; to work towards harnessing this data, we have recently expanded the set of manually generated schemas to include 17 new schemas for common dependency-management tools.}

\vspace{-0.17cm}
\section{Limitations}

Although both the challenges and techniques detailed in this paper are, in theory, applicable to a wide range of DevOps artifacts, the
dataset we provide consists solely of Dockerfiles. Futhermore, these Dockerfiles come from a single source: GitHub. It is possible that
other DevOps artifacts are not as amenable to the ideas we present. 

\vspace{-0.17cm}
\section{Summary}

DevOps artifacts in general, and Dockerfiles specifically, often see less
support than traditional program artifacts in terms of Interactive Development
Environment (IDE) extensions and tooling. We offer a large dataset of
Dockerfiles, in five different representations, to bootstrap research in the
realm of better developer assistance for DevOps and Docker. As part of these
datasets, we also contribute tools geared towards addressing some of the
challenges associated with DevOps artifacts. Namely, we provide tools to perform
various levels of parsing to uncover structure within the nested languages
present in many DevOps artifacts.

\vspace{-0.17cm}
\begin{acks}
  \added[id=J]{Supported, in part, by a gift from \grantsponsor{GS100000001}{Rajiv and Ritu Batra}{} and
  by \grantsponsor{GS100000002}{ONR}{https://www.onr.navy.mil/} under grants~\grantnum{GS100000002}{N00014-17-1-2889} and~\grantnum{GS100000002}{N00014-19-1-2318}. 
  Any opinions, findings, and conclusions or recommendations
  expressed in this publication are those of the authors,
  and do not necessarily reflect the views of the sponsoring
  agencies.}
\end{acks}

\bibliographystyle{ACM-Reference-Format}
\bibliography{bib/docker}

\end{document}